\documentclass[final,3p,times,twocolumn]{elsarticle}
\usepackage[latin9]{inputenc}
\usepackage{amsmath}
\usepackage{graphicx}
\usepackage[unicode=true,
 bookmarks=false,
 breaklinks=false,pdfborder={0 0 1},backref=false,colorlinks=false]
 {hyperref}

\makeatletter

\providecommand{\tabularnewline}{\\}

\journal{Journal of Magnetism and Magnetic Materials}

\biboptions{sort&compress}

\makeatother

\begin{document}

\begin{frontmatter}{}

\title{Local properties of the $t$-$J$ model in a two-pole approximation
within COM}

\author{Amir Eskandari-asl$^{1}$}

\author{Adolfo Avella$^{1,2,3}$}

\address{$^{1}$Dipartimento di Fisica ``E.R, Caianiello'', Universit� degli
Studi di Salerno, I-84084 Fisciano (SA), Italy}

\address{$^{2}$CNR-SPIN, UoS di Salerno, I-84084 Fisciano (SA), Italy}

\address{$^{3}$Unit� CNISM di Salerno, Universit� degli Studi di Salerno,
I-84084 Fisciano (SA), Italy}
\begin{abstract}
In this work, we study the $t$-$J$ model using a two-pole approximation
within the composite operator method. We choose a basis of two composite
operators -- the constrained electrons and their spin-fluctuation
dressing -- and approximate their currents in order to compute the
corresponding Green's functions. We exploit the algebraic constraints
obeyed by the basis operators to close a set of self-consistent equations
that is numerically solved. This allows to determine the physical
parameters of the system such as the spin-spin correlation function
and the kinetic energy. Our results are compared to those of an exact
numerical method on a finite system to asses their reliability. Indeed,
a very good agreement is achieved through a far less numerically demanding
and a more versatile procedure. We show that by increasing the hole
doping, anti-ferromagnetic correlations are replaced by ferromagnetic
ones. The behavior on changing temperature and exchange integral is
also studied and reported.
\end{abstract}

\end{frontmatter}{}

\section{Introduction}

Strongly correlated systems have undergone extensive study for several
decades \citep{avella2012scs,avella2013scs,avella2015scs}. For lattice
systems, the onsite interaction is the most important one, a feature
which is well reflected in the Hubbard model \citep{Hubbard_all,anderson1987,RevModPhys.68.13}.
Despite its seemingly simple structure, the Hubbard model and its
extensions and derivations have been very successful in studying strongly
correlated systems \citep{montorsi1992,essler2005,tasaki1998,baeriswyl2013}.
In the strong electron-electron repulsion regime, it is possible to
derive an effective model from the Hubbard model in which the double-occupancy
is discarded according to its extreme energy cost: the so called $t$-$J$
model \citep{chao1977,chao1978,PhysRevB.37.9753,stein1997,PhysRevB.70.245124,spalek2015}.
The Hubbard and the $t$-$J$ models have been used to theoretically
understand and describe so many interesting phenomena such as Mott-Hubbard
metal-insulator transition \citep{imada1998}, non-Fermi-liquid normal
phases and high-temperature superconductivity \citep{avella1998overdoped,avella2001two,chen2016,avella2014,kloss2016,lee2006,armitage2010,hashimoto2014,chowdhury2016,tajima2016},
etc. One feature which is very interesting and still needs to be clarified
is the occurrence of a ferro-antiferromagnetic crossover in the $t$-$J$
model \citep{marder1990,hellberg1997}. It is quite well-known that,
at half filling, the system is in the anti-ferromagnetic (AF) N�el
state. Yet, Nagaoka proved that in the infinite-$U$ regime of the
Hubbard model, which corresponds to vanishing exchange integral in
the $t$-$J$ model, if we introduce one hole into the system the
ground state becomes ferromagnetic (FM) \citep{nagaoka1965,nagaoka1966,Tasaki89}.
This idea got generalized in the $t$-$J$ model by so many successive
studies which showed transition to FM phase for finite hole dopings
\citep{jayaprakash89,poilblanc1992,singh1992,putikka1992,mori1993,maska2012,bhattacharjee2018,vidmar2013two,montenegro2014}.

In studying strongly correlated systems, quasi-particles play a crucial
role. In the composite operator method (COM), the equation of motion
of the operators corresponding to the most relevant quasi-particles,
those associated with the emergent elementary excitations in the system,
are investigated. Within this method, using the properties of the
generalized Green\textquoteright s functions (GFs) of the quasi-particle
operators and their algebraic constraints, a set of self-consistent
equations are obtained from which the physical properties can be computed
\citep{avella2014,mancini2004,avella2012composite,avella2014hubbard,DiCiolo2018,odashima2005high}.
It's worth noticing that COM belongs to the large class of operatorial
approaches: the Hubbard approximations \citep{Hubbard_all}, an early
high-order GF approach \citep{kuz1977}, the projection operator method
\citep{tserkovnikov1981,tserkovnikov1982}, the works of Mori \citep{mori1965},
Rowe \citep{rowe1968}, and Roth \citep{roth1969}, the spectral density
approach \citep{nolting1989}, the works of Barabanov \citep{barabanov2001},
Val\textquoteright kov \citep{val2005}, and Plakida \citep{plakida1999,plakida2001,plakida2003,plakida2007,plakida2010},
and the cluster perturbation theory in the Hubbard-operator representation
\citep{ovchinnikov2011}.

In this work, we consider a two-pole approximation for the $t$-$J$
model on a two-dimensional lattice and focus on the quasi-particles
describing the constrained electrons and their spin-fluctuation dressing.
After computing the currents of our basis operators, we apply a generalized
mean-field approximation to project them back on the basis. These
currents, together with the integrated spectral weights of the basis,
can be exploited to get a set of self-consistent equations that allow
to compute the relevant GFs. The remaining unknowns can be related
to the GFs using the algebraic constraints obeyed by the composite
operators. The solutions of these equations reveal the physical properties
of the system in different parametric regimes. The quality of the
approximation is assessed by comparing our results to those of an
exact numerical study. We find a very good agreement while our method
is, on one hand, numerically less demanding and, on the other hand,
more versatile as it can be generalized to study several different
systems and parameter regimes. We show that the system features AF
correlations of the N�el type near half filling, while by increasing
the hole doping it develops FM ones. At higher temperatures, both
AF and FM correlations are suppressed and the paramagnetic phase is
the favored one. Moreover, as expected, higher and higher values of
the exchange integral favor AF correlations and higher values of doping
are needed for the emergence of FM fluctuations.

The article is organized as follows. In Sec.~\ref{sec:mm}, we introduce
the model and the basis operators we have chosen and describe our
method. In Sec.~\ref{sec:Results}, we present our numerical results,
assess them by comparison to exact numerical ones, and discuss their
relevant features. Finally, in Sec.~\ref{sec:Summary}, we give our
conclusions.

\section{Model and Method\label{sec:mm}}

The $t$-$J$ Hamiltonian is derived in the strongly correlated regime
of the Hubbard model ($t\ll U$) where an exchange integral $J=4t^{2}/U$
emerges \citep{chao1977,chao1978,spalek2015}. Its explicit form for
a two-dimensional lattice is given by
\begin{align}
\mathcal{H} & =-4t\sum_{i}\xi^{\dagger}\left(i\right)\cdot\xi^{\alpha}\left(i\right)-\mu\sum_{i}\nu\left(i\right)\nonumber \\
 & +\frac{J}{2}\sum_{i}\left[\nu_{k}\left(i\right)\nu_{k}^{\alpha}\left(i\right)-\nu\left(i\right)\nu^{\alpha}\left(i\right)\right],
\end{align}
where $t$, $J$, and $\mu$ are the nearest-neighbor hopping integral,
the exchange integral and the chemical potential, respectively. We
set $t$ as energy unit. In this model, double occupancy of sites
is prohibited, accordingly one has to use the operator $\xi_{\sigma}\left(i\right)=\left(1-n_{\bar{\sigma}}\left(i\right)\right)c_{\sigma}\left(i\right)$,
which describes the transition between empty and singly-occupied sites,
with $c_{\sigma}\left(i\right)$ being the annihilation operator of
an electron with spin $\sigma$ on the site $i$, and $n_{\sigma}\left(i\right)=c_{\sigma}^{\dagger}\left(i\right)c_{\sigma}\left(i\right)$.
We use the spinorial notation $\xi^{\dagger}\left(i\right)=\left(\xi_{\uparrow}^{\dagger}\left(i\right),\xi_{\downarrow}^{\dagger}\left(i\right)\right)$
and define (spin) inner product between operators: $\nu\left(i\right)=\xi^{\dagger}\left(i\right)\cdot\xi\left(i\right)=\sum_{\sigma}\xi_{\sigma}^{\dagger}\left(i\right)\xi_{\sigma}\left(i\right)$
and $\nu_{k}\left(i\right)=\xi^{\dagger}\left(i\right)\cdot\sigma_{k}\cdot\xi\left(i\right)$
are the charge and spin density operators on site $i$, respectively,
with $\sigma_{k}$ being the Pauli matrices for $k=1,2,3$. Moreover,
for every operator $\phi\left(i\right)$, its projection on the nearest
neighbor sites ($\delta_{\left\langle ij\right\rangle }$) is given
by $\phi^{\alpha}\left(i\right)=\sum_{j}\alpha_{ij}\phi\left(j\right)=\frac{1}{4}\left[\phi\left(i_{x}+1,i_{y}\right)+\phi\left(i_{x}-1,i_{y}\right)+\phi\left(i_{x},i_{y}+1\right)+\phi\left(i_{x},i_{y}-1\right)\right]$,
where for a two-dimensional square lattice $\alpha_{ij}=\frac{1}{4}\delta_{\left\langle ij\right\rangle }$.
With some straightforward calculations, one can rewrite the Hamiltonian
as \citep{DiCiolo2019}
\begin{equation}
\mathcal{H}=\sum_{i}\xi^{\dagger}\left(i\right)\cdot\left[-4t\xi^{\alpha}\left(i\right)+J\left(\widetilde{\xi}_{0}\left(i\right)+\widetilde{\xi}_{s}\left(i\right)\right)-\left(\mu+\frac{J}{2}\right)\xi\left(i\right)\right],
\end{equation}
where the operators
\begin{equation}
\widetilde{\xi}_{0}\left(i\right)=\frac{1}{2}\left(1-\nu^{\alpha}\left(i\right)\right)\xi\left(i\right),
\end{equation}
\begin{equation}
\widetilde{\xi}_{s}\left(i\right)=\frac{1}{2}\nu_{k}^{\alpha}\left(i\right)\sigma_{k}\cdot\xi\left(i\right),
\end{equation}
describe constrained electronic transitions dressed by nearest-neighbor
charge and spin fluctuations, respectively. As our basis operators
we choose the following set of operators
\begin{equation}
\boldsymbol{\psi}\left(i\right)=\left(\begin{array}{c}
\xi\left(i\right)\\
\widetilde{\xi}_{s}\left(i\right)
\end{array}\right),
\end{equation}
which reflects the fact that near half filling, the spin fluctuations
play the most important role as the system has a clear tendency towards
the AF N�el state.

In the Heisenberg picture, the current of the basis operators is
\begin{equation}
\boldsymbol{J}\left(i\right)=i\frac{\partial}{\partial t_{i}}\boldsymbol{\psi}\left(i\right)=\left[\boldsymbol{\psi}\left(i\right),\mathcal{H}\right],
\end{equation}
where we set $\hbar=1$. One can show that
\begin{alignat}{1}
J_{\xi}= & -2t\left(\xi^{\alpha}\left(i\right)+2\xi_{0}\left(i\right)+2\xi_{s}\left(i\right)\right)\nonumber \\
 & +2J\left(\widetilde{\xi}_{0}\left(i\right)+\widetilde{\xi}_{s}\left(i\right)\right)-\left(\mu+J\right)\xi\left(i\right),
\end{alignat}
\begin{align}
J_{\widetilde{\xi}_{s}}= & -t\nu_{k}^{\alpha}\left(i\right)\sigma_{k}\cdot\left(\xi^{\alpha}\left(i\right)+2\xi_{0}\left(i\right)+2\xi_{s}\left(i\right)\right)\nonumber \\
 & +\left(-2t\left[\xi^{\dagger}\left(i\right)\cdot\sigma_{k}\cdot\xi^{\alpha}\left(i\right)\right]^{\alpha}+2t\left[\xi^{\dagger\alpha}\left(i\right)\cdot\sigma_{k}\cdot\xi\left(i\right)\right]^{\alpha}\right)\sigma_{k}\cdot\xi\left(i\right)\nonumber \\
 & -\mu\widetilde{\xi}_{s}\left(i\right)-\frac{J}{2}\nu_{k}^{\alpha}\left(i\right)\nu^{\alpha}\left(i\right)\sigma_{k}\cdot\xi\left(i\right)\nonumber \\
 & +\frac{J}{2}\nu_{k}^{\alpha}\left(i\right)\nu_{g}^{\alpha}\left(i\right)\sigma_{k}\cdot\sigma_{g}\cdot\xi\left(i\right)+Ji\epsilon_{kgh}\left[\nu_{g}^{\alpha}\left(i\right)\nu_{h}\left(i\right)\right]^{\alpha}\sigma_{k}\cdot\xi\left(i\right),
\end{align}
where $\xi_{0}\left(i\right)=\frac{1}{2}\left(1-\nu\left(i\right)\right)\xi^{\alpha}\left(i\right)$
and $\xi_{s}\left(i\right)=\frac{1}{2}\nu_{k}\left(i\right)\sigma_{k}\cdot\xi^{\alpha}\left(i\right)$
can be considered as counterparts of $\widetilde{\xi}_{0}\left(i\right)$
and $\widetilde{\xi}_{s}\left(i\right)$, respectively. Taking into
account only nearest neighbor contributions, these latter higher-order
operators can be approximated as
\begin{align}
\xi_{0}\left(i\right)\simeq & 4\widetilde{\xi}_{0}^{\alpha}\left(i\right)-\frac{3}{2}\left(1-\nu\right)\xi^{\alpha}\left(i\right),
\end{align}
\begin{align}
\xi_{s}\left(i\right)\simeq & 4\widetilde{\xi}_{s}^{\alpha}\left(i\right).
\end{align}
Moreover, one can approximate $\widetilde{\xi}_{0}\left(i\right)$
by projecting it on $\xi\left(i\right)$ as \citep{DiCiolo2019}
\begin{equation}
\widetilde{\xi}_{0}\left(i\right)\approx\frac{2-3\nu+\chi_{c}^{\alpha}}{4\left(1-\frac{\nu}{2}\right)}\xi\left(i\right)-\frac{C_{11}^{\alpha}}{2\left(1-\frac{\nu}{2}\right)}\xi^{\alpha}\left(i\right).
\end{equation}
Finally, considering a paramagnetic and homogenous phase and using
a mean-field like approximation we can write the currents in the form
\citep{mancini2004,avella2012composite,avella2014hubbard,avella2014}
\begin{equation}
J_{a}\left(i\right)=\sum_{j}\sum_{b}\varepsilon_{ab}\left(i,j\right)\psi_{b}\left(j\right).
\end{equation}
The Fourier transform of the $\boldsymbol{\varepsilon}$ matrix reads
as
\begin{align}
\varepsilon_{11}\left(\boldsymbol{k}\right)= & 16t\frac{C_{11}^{\alpha}}{2-\nu}\alpha^{2}\left(\boldsymbol{k}\right)\nonumber \\
 & +\left(6t\left(\frac{2}{3}-\nu\right)-8t\frac{2-3\nu+\chi_{c}^{\alpha}}{2-\nu}-2J\frac{C_{11}^{\alpha}}{2-\nu}\right)\alpha\left(\boldsymbol{k}\right)\nonumber \\
 & +J\frac{2-3\nu+\chi_{c}^{\alpha}}{2-\nu}-\mu-J,\\
\varepsilon_{12}\left(\boldsymbol{k}\right)= & -16t\alpha\left(\boldsymbol{k}\right)+2J,\\
\varepsilon_{21}\left(\boldsymbol{k}\right)= & -6tC_{11}^{\alpha}\left(1+\frac{1}{2-\nu}\right)\alpha^{2}\left(\boldsymbol{k}\right)\nonumber \\
 & +\biggr(-\frac{3}{4}t-t\frac{9}{4}\chi_{s}^{\alpha}+6tC_{11}^{\alpha^{2}}-\frac{15}{4}t\left(1-\nu\right)\nonumber \\
 & +6t\frac{2-3\nu+\chi_{c}^{\alpha}}{4-2\nu}+3J\frac{C_{11}^{\alpha}}{8-4\nu}\biggr)\alpha\left(\boldsymbol{k}\right)\nonumber \\
 & +\frac{3J}{8}+\frac{3}{2}tC_{11}^{\alpha}-\frac{3J}{4}\frac{2-3\nu+\chi_{c}^{\alpha}}{4-2\nu},\\
\varepsilon{}_{22}\left(\boldsymbol{k}\right)= & 2t\alpha\left(\boldsymbol{k}\right)-\left(\mu+\frac{3J}{4}+\frac{3J}{4}\nu\right),
\end{align}
where $\nu=\left\langle \nu\left(i\right)\right\rangle $ is the average
electron number per site which can vary between 0 and 1 (half filling)
depending on the doping. $C_{ab}^{\alpha^{n}}=\left\langle \psi_{a}^{\alpha^{n}}\left(i\right)\psi_{b}^{\dagger}\left(i\right)\right\rangle $
is the generalized correlation matrix {[}$\phi^{\alpha^{n}}\left(i\right)=\sum_{j}\alpha_{ij}^{n}\phi\left(j\right)${]}
with $n$ being a non-negative integer: $\alpha_{ij}^{0}=\delta_{ij}$,
$\alpha_{ij}^{1}=\alpha_{ij}$, and for $n>1$, $\alpha_{ij}^{n}=\sum_{l_{1},..,l_{n-1}}\alpha_{il_{1}}\alpha_{l_{1}l_{2}}...\alpha_{l_{n-1}j}$.
$\chi_{c}^{\alpha}=\left\langle \nu\left(i\right)\nu^{\alpha}\left(i\right)\right\rangle $
and $\chi_{s}^{\alpha}=\frac{1}{3}\sum_{k=1}^{3}\left\langle \nu_{k}\left(i\right)\nu_{k}^{\alpha}\left(i\right)\right\rangle $
are the charge-charge and spin-spin correlation functions, respectively.

The normalization matrix of the basis operators is defined as
\begin{equation}
I_{ab}\left(i,j\right)=\left\langle \left\{ \psi_{a}\left(i\right),\psi_{b}^{\dagger}\left(j\right)\right\} \right\rangle .
\end{equation}
Once again, we use mean-field-like approximations and perform Fourier
transformation to obtain
\begin{align}
I_{11}\left(\boldsymbol{k}\right)= & 1-\frac{1}{2}\nu,\\
I_{12}\left(\boldsymbol{k}\right)= & \frac{3}{4}\chi_{s}^{\alpha}+\frac{3}{2}\alpha\left(\boldsymbol{k}\right)C_{11}^{\alpha},\\
I_{22}\left(\boldsymbol{k}\right)= & \frac{3}{16}\left(-\frac{1}{2}\chi_{c}^{\alpha}-\chi_{s}^{\alpha}+\nu\right)-\alpha\left(\boldsymbol{k}\right)C_{12}^{\alpha}\nonumber \\
 & +\frac{3\alpha\left(\boldsymbol{k}\right)}{16}C_{11}^{\alpha}+\left(2\alpha^{2}\left(\boldsymbol{k}\right)-\frac{1}{2}\right)\frac{4}{3}C_{12}^{\alpha^{2}},
\end{align}
where in the last line we used the so called spherical approximation
\citep{avella2003bosonic,mancini2004}.

In order to obtain the self-consistent set of equations, we define
the retarded GF as follow.
\begin{equation}
G_{ab}^{R}\left(i,j\right)=\theta\left(t_{i}-t_{j}\right)\left\langle \left\{ \psi_{a}\left(i\right),\psi_{b}^{\dagger}\left(j\right)\right\} \right\rangle ,
\end{equation}
where $i$ stands for both time and site indices. For a basis of $n$
operators, GF is an $n\times n$ matrix (in our case $n=2$). Then,
using the current equations and performing Fourier transformation
in space and time one can show
\[
\boldsymbol{G}^{R}\left(\boldsymbol{k},\omega\right)=\left(\omega-\boldsymbol{\varepsilon}\left(\boldsymbol{k}\right)\right)^{-1}\boldsymbol{I}\left(\boldsymbol{k}\right),
\]
which results in the following explicit form
\begin{equation}
\boldsymbol{G}^{R}\left(\boldsymbol{k},\omega\right)=\sum_{m=1}^{n}\frac{\boldsymbol{\sigma}^{\left(m\right)}\left(\boldsymbol{k}\right)}{\omega-\omega_{m}\left(\boldsymbol{k}\right)+i0^{+}},\label{eq:GFR}
\end{equation}
where $\omega_{m}\left(\boldsymbol{k}\right)$ is the m-th eigenvalue
of $\boldsymbol{\varepsilon}\left(\boldsymbol{k}\right)$, and 
\begin{equation}
\sigma_{ab}^{\left(m\right)}\left(\boldsymbol{k}\right)=\Omega_{am}\left(\boldsymbol{k}\right)\sum_{c=1}^{2}\Omega_{mc}^{-1}\left(\boldsymbol{k}\right)I_{cb}\left(\boldsymbol{k}\right),\label{eq:sigma}
\end{equation}
in which $\boldsymbol{\Omega}\left(\boldsymbol{k}\right)$ is an $n\times n$
matrix whose columns are the eigenvectors of $\boldsymbol{\varepsilon}\left(\boldsymbol{k}\right)$.
Using Eq,~\ref{eq:GFR}, one can obtain a generalized form of the
fluctuation-dissipation theorem as \citep{mancini2004} 
\begin{equation}
\boldsymbol{C}\left(\boldsymbol{k},\omega\right)=2\pi\sum_{m=1}^{n}\left[1-f_{F}\left(\omega_{m}\left(\boldsymbol{k}\right)\right)\right]\boldsymbol{\sigma}^{\left(m\right)}\left(\boldsymbol{k}\right)\delta\left(\omega-\omega_{m}\left(\boldsymbol{k}\right)\right),
\end{equation}
where $f_{F}$ is the Fermi distribution function. Performing the
inverse Fourier transformation, we obtain
\begin{align}
C_{ab}^{\kappa}= & \frac{2\pi}{N}\sum_{\boldsymbol{k}}\kappa\left(-\boldsymbol{k}\right)\sum_{l=1}^{n}\left[1-f_{F}\left(\omega_{l}\left(\boldsymbol{k}\right)\right)\right]\sigma_{ab}^{\left(l\right)}\left(\boldsymbol{k}\right),\label{eq:cxab}
\end{align}
where $\kappa$ can be any lattice projection operator. This relation
shows how the self-consistent procedure works. For calculating the
GFs, we need $\boldsymbol{I}\left(\boldsymbol{k}\right)$ and $\boldsymbol{\varepsilon}\left(\boldsymbol{k}\right)$,
which are determined by the correlation functions. On the other hand,
the correlation functions are determined by the GFs through the fluctuation-dissipation
theorem, Eq.~\ref{eq:cxab}. In order to close the set of self-consistent
equations, we use the following algebraic constraints obeyed by the
basis operators.
\begin{align}
C_{11}^{\delta}= & 1-\nu,
\end{align}
\begin{equation}
C_{12}^{\delta}=0,
\end{equation}
\begin{alignat}{1}
C_{22}^{\delta}= & -\frac{3}{16}\chi_{c}^{\alpha}+\frac{3}{16}\nu.
\end{alignat}
Having a closed set of self-consistent equations, we numerically solve
it to obtain the physical properties of the system.

\begin{figure}[t!]
\noindent \centering{}%
\begin{tabular}{c}
\includegraphics[width=0.98\columnwidth]{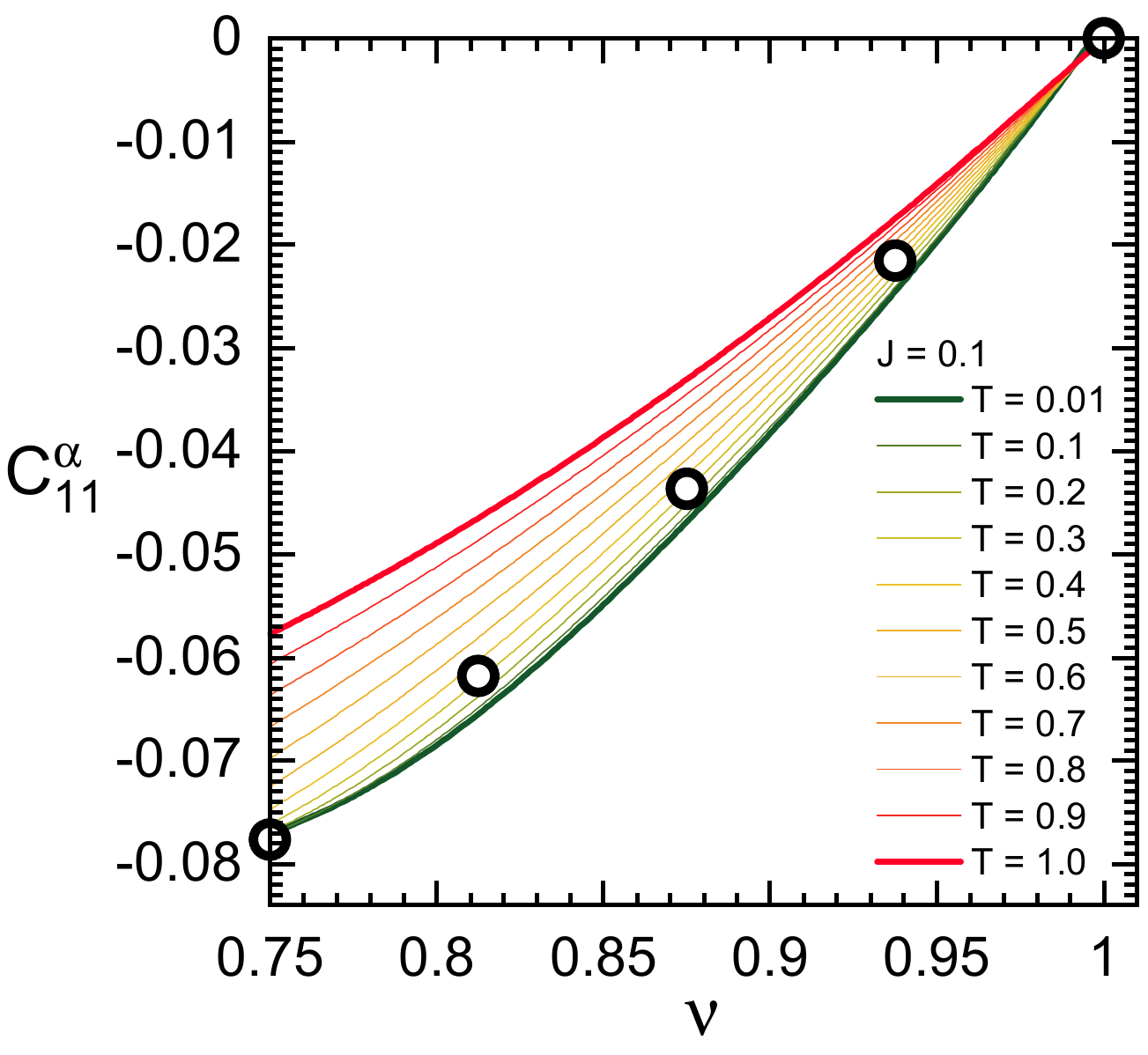}\tabularnewline
\end{tabular}\caption{$C_{11}^{\alpha}=\left\langle \xi^{\alpha}\left(i\right)\xi^{\dagger}\left(i\right)\right\rangle $
, as a function of filling, $\nu$, for $J=0.1$ and temperature T
ranging from 0.01 to 1. Circles are ED data extracted from Ref.~\citep{dagotto1992}.
$C_{11}^{\alpha}$ is proportional to the kinetic energy, $K=8tC_{11}^{\alpha}$.\label{fig:1_ca11}}
\end{figure}

\section{Results\label{sec:Results}}

In this section, we present our numerical results. In Fig.~\ref{fig:1_ca11},
we show $C_{11}^{\alpha}=\left\langle \xi^{\alpha}\left(i\right)\xi^{\dagger}\left(i\right)\right\rangle $
as a function of electron density per site, $\nu$, for $J=0.1$ and
temperature $T$ ranging from 0.01 to 1. The circles are numerical
data extracted from Ref.~\citep{dagotto1992} and correspond to exact
diagonalization (ED) results at zero temperature for a finite cluster.
There is a clear agreement with our results although we need a small
finite temperature to compensate for the finite size effects. At half
filling ($\nu=1$), $C_{11}^{\alpha}$ vanishes, as it is proportional
to the kinetic energy by the relation $K=8tC_{11}^{\alpha}$. Since
each site is occupied exactly by one electron there is no possibility
for electrons to move, and kinetic energy vanishes. Our results show
that the kinetic energy decreases by increasing the temperature, which
means the thermally excited states of the system do not favor hole
mobility, as it will be clarified in the following.

\begin{figure}[t!]
\noindent \centering{}%
\begin{tabular}{c}
\includegraphics[width=0.98\columnwidth]{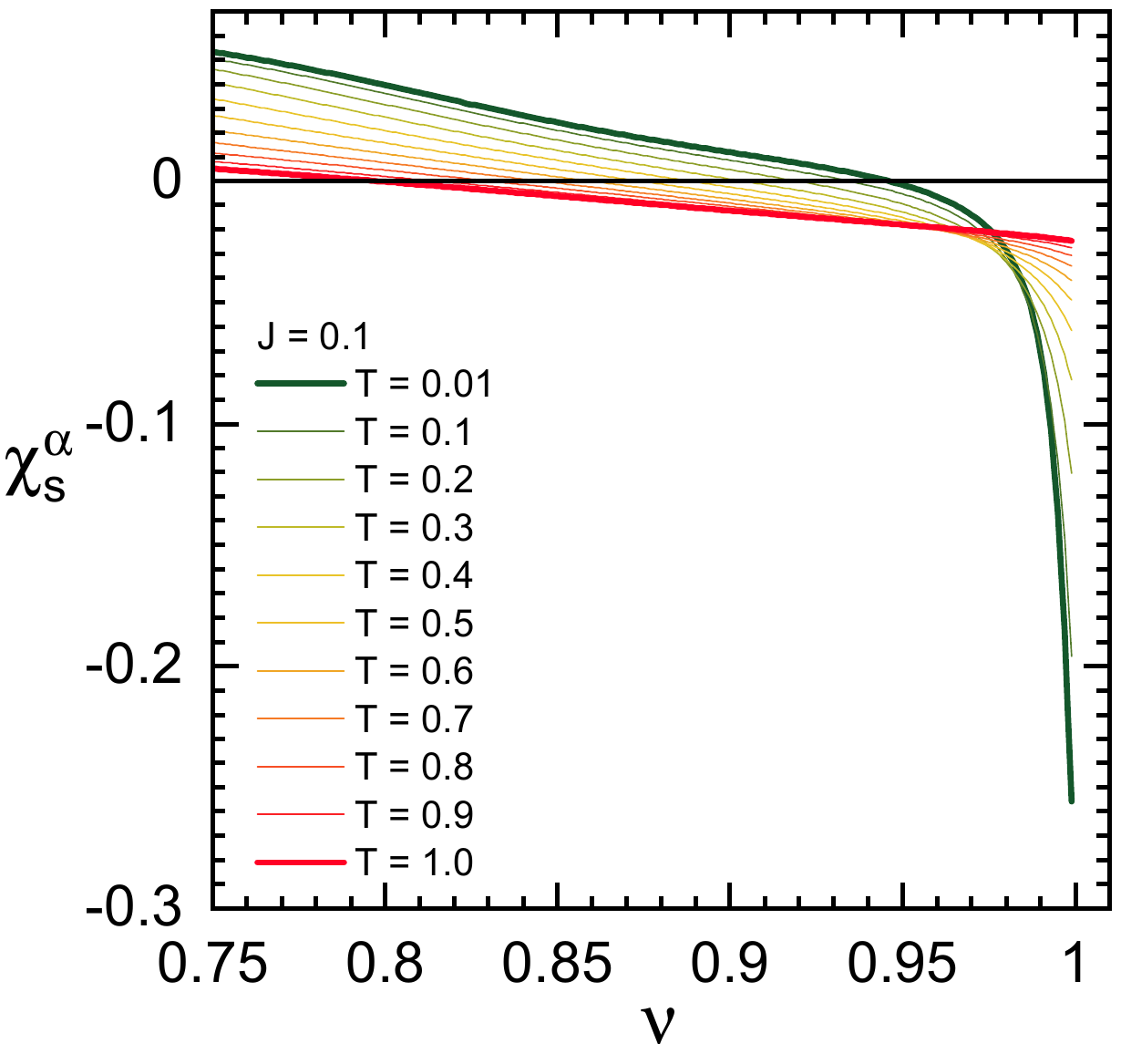}\tabularnewline
\includegraphics[width=0.98\columnwidth]{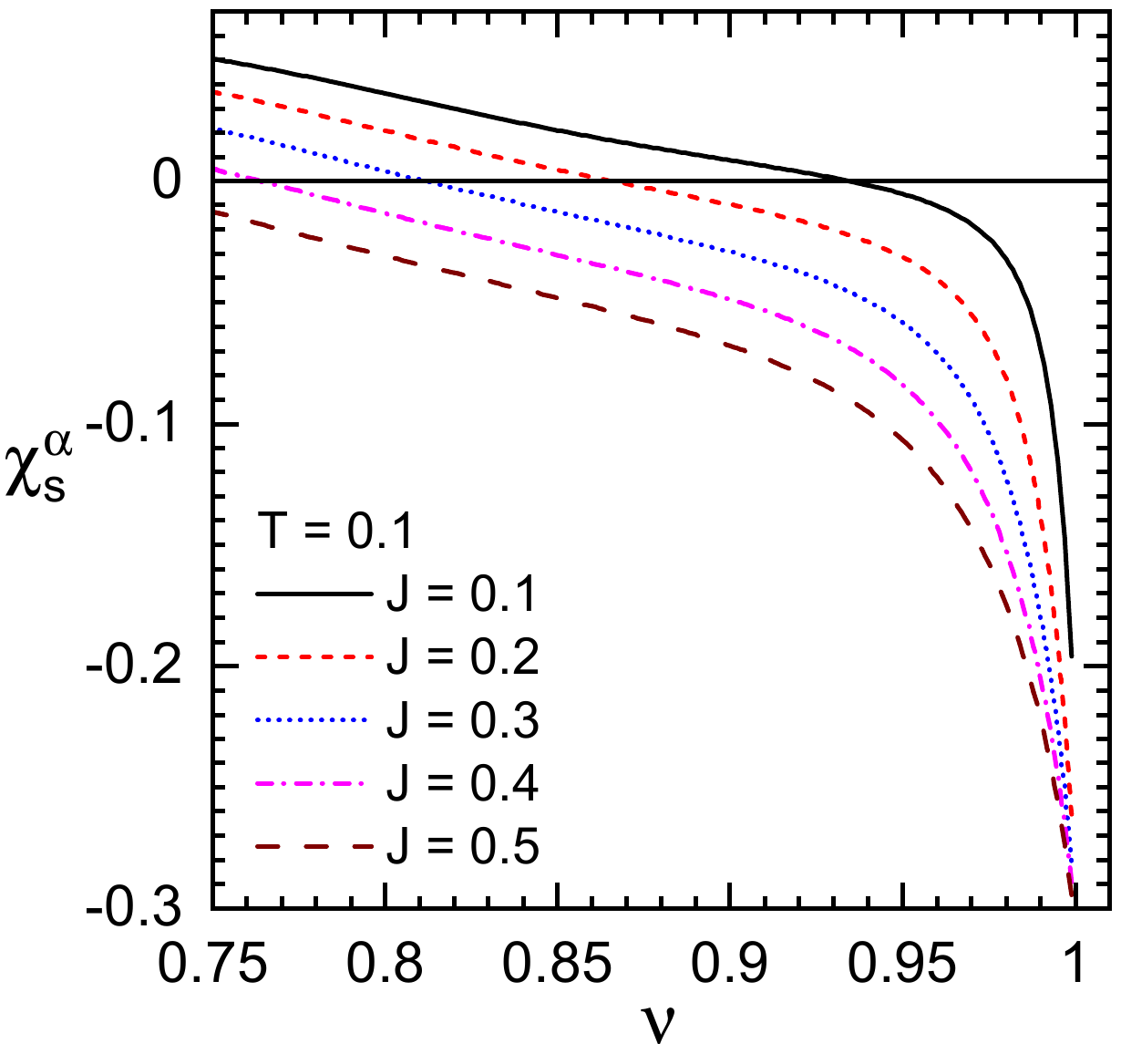}\tabularnewline
\end{tabular}\caption{$\chi_{s}^{\alpha}$ as a function of $\nu$: (top) same parameters
as Fig.~\ref{fig:1_ca11}; (bottom) $T=0.1$ and $J$ ranging from
0.1 to 0.5.\label{fig:2_chis}}
\end{figure}

Although we considered a paramagnetic phase, we can still investigate
the tendency of the system towards other (ordered) phases. In Fig.~\ref{fig:2_chis},
top panel, we plot the spin-spin correlation function, $\chi_{s}^{\alpha}$,
as a function of $\nu$, with same parameters as Fig.~\ref{fig:1_ca11}.
For low enough temperatures, we have AF correlations near half filling.
This clearly shows that our solution correctly captures the behavior
in this regime, consistently with the well-established AF N�el state
at half filling. The FM phase in the $t$-$J$ model has been predicted
in the literature \citep{jayaprakash89,marder1990,poilblanc1992,singh1992,putikka1992,mori1993,hellberg1997,maska2012,bhattacharjee2018,vidmar2013two,montenegro2014}:
mobile holes can form Nagaoka polarons which results in a FM ordering
\citep{maska2012,vidmar2013two}. We witness a similar behavior here,
i.e., once enough holes are present in the system, FM correlations
clearly emerge and they overcome the AF ones, whose correlation lengths
decrease rapidly with doping \citep{singh1992}. Increasing the temperature
results in weakening of both AF and FM correlations: the paramagnetic
phase becomes the most favorable one. Let us now come back to the
decrease of the kinetic energy on increasing the temperature reported
in Fig.~\ref{fig:1_ca11}. This behavior has different explanations
in different doping regimes. Near half filling, the AF correlations
get weaker and weaker on increasing the temperature, inhibiting the
virtual hopping processes because of the Pauli exclusion principle.
Accordingly, the kinetic energy decreases. On the other hand, at intermediate
fillings, significant FM correlations result from the formation of
Nagaoka polarons, which requires mobile holes and induce a gain in
kinetic energy. By increasing the temperature, the FM correlations
too become weaker and weaker and, consequently, the kinetic energy
decreases also in this case.

In Fig.~\ref{fig:2_chis} bottom panel, we plot the spin-spin correlation
function as a function of $\nu$ for $T=0.1$ and $J$ ranging from
0.1 to 0.5. For larger and larger values of $J$: $\left(i\right)$
the AF correlations increase, which shows a stronger tendency towards
AF for larger exchange integrals, as expected; (ii) the emergence
of FM correlations requires larger and larger values of doping in
order to overcome the stronger and stronger AF correlations.

\section{Summary\label{sec:Summary}}

In summary, we performed a two-pole study of the $t$-$J$ model within
COM. In our calculations, we considered the constrained electrons
and their spin dressing as fundamental quasi particles. By exploiting
mean-field-like approximations, we projected back the operatorial
currents on the basis operators. We used similar approximations to
calculate the normalization matrix within COM. These relations can
be combined with the algebraic constraints obeyed by the operators
to give a closed set of self-consistent equations which can be numerically
solved.

Our results for the kinetic energy are in a very good agreement with
those of ED for finite clusters, while our method is numerically less
demanding and also more versatile. Moreover, we show that the system
undergoes a smooth transition between small and intermediate doping
regimes where it features AF and FM correlations, respectively. By
increasing the temperature, both AF and FM correlations are weakened
and, consequently, the kinetic energy decreases due to the inhibition
of exchange virtual processes and polaron formation, respectively.
Increasing the exchange integral strengthens the AF fluctuations,
as expected, and forces the FM fluctuations to emerge at higher values
of doping.

\section*{Acknowledgments}

Authors acknowledge support by MIUR under Project No. PRIN 2017RKWTMY.

\bibliographystyle{elsarticle-num}
\bibliography{biblio}

\end{document}